\documentstyle [12pt,a4,subeqn] {article}
\begin{document}
\thispagestyle{empty}
\begin{center}
{\Large	\bf Chaos in the Quantum Double Well Oscillator:\\
The Ehrenfest View Revisited} \\
\vspace{.6in}
{\bf J.K. Bhattacharjee} \\
\vspace{0.1in}
{\sl Department of Theoretical Physics} \\
{\sl Indian Association for the Cultivation of Science} \\
{\sl Kolkata 700 032, India.}  \\
\vspace{0.4in}
{\bf Analabha Roy} \\
\vspace{0.1in}
{\sl Department of Physics} \\
{\sl Indian Institute of Technology} \\
{\sl Kanpur 208 016, India.} \\
\vspace{.8in}
\end{center}
\begin{abstract}
We treat the double well quantum oscillator from the standpoint of the 
Ehrenfest equation but in a manner different from Pattanayak and
Schieve. We show that for short times there can be chaotic motion due
to quantum fluctuations, but over sufficiently long times the
behaviour is normal. 
\end{abstract}

\newpage
\baselineskip 24pt

It is generally agreed that the full quantum dynamics does not exhibit 
chaos. For systems which exhibit chaotic dynamics in the classical
limit, it was clearly established by Fishman, Grempel and Prange$^1$
that there exists a critical time $t \sim 0$ ($1/h^{1/r}$) beyond
which the dynamics crosses over to the quantal behaviour. The exponent 
`$r$' was found to be 6.039 for systems which showed period-doubling
bifurcations in the classical limit and 3.04 for the disappearance of
the final KAM trajectory in the standard map. It was conjectured in
the late eighties that there could be systems where the classical
dynamics is obviously regular but the semiquantal dynamics can be
chaotic (for a more precise explanation of the term ``semiquantal'',
c.f. Pattanayak and Schieve$^2$). In support of this conjecture,
Pattanayak and Schieve$^2$ explored the semiquantal dynamics of the
double well oscillator governed by the Hamiltonian
$$
H = {P^2 \over 2} - {1 \over 2} \ x^2 + {\lambda \over 4} \ x^4
$$

The classical dynamics of this oscillator is obviously regular, as
explained in Landau-Lifshitz (Vol. 1), being a periodic trajectory
centered about $x  
= \pm {1\over \sqrt{\lambda}}$ for total energy $E$ in the range $0 > E 
> -1/4$ and about $x=0$ for $E>0$. It was shown, based on an
Ehrenfest equation approach that in the semiquantal limit the quantum
fluctuations cause the dynamics of this oscillator to become chaotic
with four repelling zones in the phase space. Now, the full quantum
dynamics of this oscillator should be regular just as all other fully
quantum dynamics. Hence, we believe that the dynamics of this
oscillator should cross over from a chaotic dynamics to a regular
dynamics as one goes from a semiquantal to a fully quantum limit. The
crossover should be characterized by a time $t_0$. Unlike the cases
treated by Grempel et al$^1$, we believe that the time scale here
should be exponentially big. By using the Ehrenfest formulation, we
will take the use of it by Pattanayak and Schieve$^2$ from a different 
standpoint. The Ehrenfest dynamics of the centroid $<x>$ of a wave
packet is given by
$$
<\ddot x> = <x> - \lambda <x^3> ,
$$
i.e.
\begin{equation}
<\ddot x> - <x> + \lambda <x>^3 = \lambda [<x>^3 - <x^3>] = Q
\end{equation}

Had the centroid followed the classical trajectory, the equations of
motion would have been given by Eq. (1) with $Q=0$. $Q$ represents the 
effect of quantum fluctuations and acts as a drive for the motion of
the centroid of the wave packet. The effect of quantum mechanics on
the dynamics is through the term $Q$. For short times, this drive is
sinusoidal in time and that is when the dynamics is like that of a
double well oscillator (Duffing Oscillator) in the presence of an
oscillating field. This dynamics is known to be chaotic. At long times 
the drive changes character due to quantum interference effects and
that is when the chaos induced by quantum fluctuation should go away.

In order to make the above point in a clearer fashion, we work with an 
average energy that is close to the ground state of the double
well. With this deep an energy, tunneling has a low probability and
there are two scales to the problem; one in which a wave packet
oscillates inside a well and another in which it can tunnel from one
well to another. With this in mind, the class of wave packets that we
will work with is
\begin{equation}
\psi (x,t) = N_1 (t) e^{- (x-a_0 - \epsilon (t))^2/2 b^2_0} + N_2 (t) 
e^{- (x+a_0 + \epsilon (t))^2/2 b^2_0} .
\end{equation}

We choose $a_0$ and $b_0$ such that with $\epsilon = 0$ and $N_1 = \pm 
N_2, \psi (x,t)$ yields the ground state and the first excited state
in the variational sense. With $b_0 \ll a_0$ and to the lowest order
in $b^{-1}_0$ and $e^{-(a_0/b_0)^2}$, we have 
\begin{equation}
a^2_0 + {3\over 2} \ b^2_0 = {1\over \lambda} + {\hbar^2 \over
4 m b^6_0 \lambda} e^{- a^2_0/b^2_0} .
\end{equation}

Normalization yields
\begin{equation}
(N^2_1 + N^2_2 + 2N_1 N_2 e^{- (a^2_0/b^2_0)}) b_0 \sqrt{\pi} = 
1 .
\end{equation}

Writing down $<H>$ for the above state, we find that the two states
with $\epsilon = 0$ and $N_1 = \pm N_2$ are separated by $\Delta E =
\displaystyle{{a^2_0 \hbar^2 \over 2m b_0^4} e^{-(a^2_0/b^2_0)}}$ which equals
$\displaystyle{{1\over \sqrt{\lambda}} e^{- \sqrt{2m}/\lambda\hbar}}$,
when we use 
Eq. (3), keeping in mind that $h$ is small. The time of quantum
tunneling from one well to another is expected to the $\sqrt{\lambda}
e^{\sqrt{2m}/\lambda\hbar}$ and over this time scale we expect the
chaotic dynamics to be smoothened out. 

Turning to the dynamics, we now have, using Eq. (2), the result
\begin{equation}
<x> = (a_0 + \epsilon) {|N_1|^2 - |N_2|^2 \over |N_1|^2 + |N_2|^2 + 2
{\rm Re} (N_1^\ast N_2) e^{-(a_0 +\epsilon)^2 /b_0^2}}
\end{equation}
and a similar expression may be found for $<x^3>$. We incorporate
phase terms in $N_1$ and $N_2$ into the overall phase term of
$\psi$. Thus we can say without loss of generality that $\{N_1, N_2\} \
\epsilon \ \ \Re$. In keeping with our approximations, $e^{-(a_0 
+ \epsilon)^2/b^2_0}$ is small and Eq. (5) can be taken to be (see
Eq.(4) as well)
\subequations
\begin{equation}
<x> = (a_0 + \epsilon) (N_1^2 - N_2^2) , 
\end{equation}
\begin{equation}
<x^3> = (a_0 + \epsilon) \left[(a_0 + \epsilon)^2 + {3\over 2}
b^2_0\right] (N_1^2 - N^2_2) .
\end{equation}
\endsubequations

The quantum fluctuation $Q$ in Eq.(1) is given by 
\begin{eqnarray}
Q = \lambda [ <x>^3 - <x^3> ] &=& \lambda \left\{ (a_0 + \epsilon)^3
(N_1^2 - N_2^2) [(N_1^2 - N^2_2)^2 - 1]\right\} \nonumber \\ [2mm]
&& - {3\over 2} \lambda (a_0 + \epsilon) b^2_0 (N^2_1 - N_2^2) .
\end{eqnarray}
The point to note is that, at small time scales, $Q$ is a fast
oscillation (i.e. same scale as frequency oscillation of $Q=0$ set by
oscillation in $\epsilon$) and can induce$^3$ chaos in the Duffing 
Oscillator. If $E$ averages to $0$, we have $<x> \approx (N^2_1 -
N_2^2) a_0$, $<x^3> = (N^2_1 - N_2^2) a_0 (a^2_0 + {3\over 2}
b^2_0)$ and Eq.(1) reads
$$
{d^2\over dt^2} (N^2_1 - N^2_2) + \lambda a_0 \left(a^2_0 + {3\over 2}
b^2_0 - {1\over \lambda}\right) (N^2_1 - N^2_2) = 0 ,
$$
i.e.
\begin{equation}
{d^2\over dt^2} (N^2_1 - N^2_2) + {a_0 \hbar^2 \over 4m b^6_0} 
e^{- a^2_0/b^2_0} (N^2_1 - N^2_2) = 0 .
\end{equation}
It is clear from Eq.(8) that the time scale over which $N^2_1 - N^2_2$
oscillates is ${\cal O} (e^{a^2_0/b^2_0})$ which is very big. In this limit,
the quantum fluctuation $Q$ of Eq.(7) shows an oscillatory behaviour
with a long time scale. The effect of this low frequency drive is
negligible.

For shorter time scales, where $N_1^2$ is $\gg N^2_2$ and is almost
constant, we have $<x> = (a_0 + \epsilon)$ and 
$<x^3> = (a_0 + \epsilon)^3 + {3\over 2} b^2_0 (a_0 + \epsilon)$. This 
makes the dynamics of $\epsilon$ to be given by
\begin{equation}
\ddot\epsilon + 2a^2_0 \lambda \epsilon + 3a_0 \lambda \epsilon^2 + \lambda
\epsilon^3 = 0 . 
\end{equation}
Since $a^2_0 \approx {1\over \lambda}$, we have, for small $\lambda$, 
\begin{equation}
\ddot\epsilon + 2 \epsilon = 0 .
\end{equation}

The time period for the oscillations of $\epsilon$ is ${\cal O} (1)$. The
quantum fluctuation term $Q$ is now given by $Q \simeq - {3\over 2}
\lambda (a_0 + \epsilon) b^2_0$, which is a drive with periodicity
matching that of the double well with $Q=0$. This drive is capable of 
introducing chaos$^3$ and we believe that this is the phenomenon
reported by Pattanayak and Schieve$^2$. This crossover in the drive
generated by quantum fluctuation is what we wanted to describe. While
it is generally believed that quantum dynamics should be nonchaotic,
there is a short time regime where the quantum fluctuation$^{2,4}$ can 
make the centroid of a wave packet follow a classical trajectory which 
is chaotic. For sufficiently long times, the drive changes character
and we do not have a chaotic response. The time scale for this
crossover is exponentially large. It should be noted that if we had
started with $N_2 \simeq 0$, after a very long time, we would find the
system in a situation where $N_1 \approx 0$ and the chaotic phenomenon 
would return. This is another way of describing the quantum noise
induced chaotic oscillations reported recently$^5$.

\end{document}